\documentclass[useAMS,usenatbib]{mn2e}
\usepackage{graphics, graphicx,times}

\newif\ifAMStwofonts

\def\lapp{\ifmmode\stackrel{<}{_{\sim}}\else$\stackrel{<}{_{\sim}}$\fi}
\def\gapp{\ifmmode\stackrel{>}{_{\sim}}\else$\stackrel{>}{_{\sim}}$\fi}

\title[An empirical model for the beams of radio pulsars]
{An empirical model for the beams of radio pulsars}
\author[Karastergiou \& Johnston]
{Aris Karastergiou$^1$ \& Simon Johnston$^2$\\
$^1$ IRAM, 300 rue de la Piscine, Domaine Universitaire, Saint Martin d'Heres, France.\\
$^2$Australia Telescope National Facility, CSIRO, P.O. Box 76, 
Epping, NSW 1710, Australia.
}
\date{\today}
\pagerange{\pageref{firstpage}--\pageref{lastpage}}
\pubyear{2005}
\begin{document}
\maketitle
\label{firstpage}

\begin{abstract}
Motivated by recent results on the location of the radio emission in
pulsar magnetospheres, we have developed a model which can account for
the large diversity found in the average profile shapes of pulsars. At
the centre of our model lies the idea that radio emission at a
particular frequency arises from a wide range of altitudes above the
surface of the star and that it is confined to a region close to the
last open field lines.  We assert that the radial height range over
which emission occurs is responsible for the complex average pulse
shapes rather than the transverse (longitudinal) range proposed in
most current models. By implementing an abrupt change in the height
range to discriminate between young, short-period, highly-energetic
pulsars and their older counterparts, we obtain the observed
transition between the simple and complex average pulse profiles
observed in each group respectively. Monte Carlo simulations are used
to demonstrate the match of our model to real observations.
\end{abstract}


\section{Introduction}
In this paper, we create an empirical model for the beams of radio
pulsars, a model which we believe can account for much of the
phenomenology seen in the observations.  We use the model to generate
a large number of artificial pulsar profiles and statistically compare
them to recently acquired, high quality pulsar data. We believe the
simplicity of the model to be one of its main strengths, in particular
for a better theoretical understanding of the radio emission process.

The integrated profiles of radio pulsars are one of their defining
characteristics. Made up from the summation of several thousand single
pulses, in the vast majority of cases these integrated profiles are
highly stable over decades of observing. It is precisely this profile
stability that allows for the determination of highly accurate pulse
arrival times leading to the exquisite tests of physics for which
pulsar astronomy is renowned (e.g. Kramer et al. 2006\nocite{ksm+06}).

Each pulsar has its own unique integrated profile consisting of a
(usually small) number of Gaussian-shaped components (Kramer et
al. 1994\nocite{kwj+94}), and even a cursory examination of the many
hundreds of profiles available in the literature shows that they come
in a bewildering variety of forms. This variety owes partly to the
shape of the pulsar beam and partly to the geometry of the star and
the relative orientation to the observer. Unfortunately, it is very
difficult to determine the viewing geometry, and therefore also the
true pulsar beam shape. Nevertheless, over the years many different
attempts have been made to classify the integrated profiles into
groups with well defined characteristics.  Two different ideas have
emerged, each with their own pros and cons.  In the work of Rankin and
co-workers (Rankin 1983; Rankin 1993; Mitra \& Rankin
2002)\nocite{ran83,ran93,mr02a}, emission is recognised as arising
from near the magnetic pole of the star (`core' emission) and in
concentric rings around the pole (`cone' emission). This is a natural
explanation for symmetric pulse profiles and for profiles with an odd
number of components. In contrast, Lyne \& Manchester
(1988)\nocite{lm88} argue that the emission cone is patchy with
emission occurring at random locations in the beam convolved with an
annular `window function'. This more obviously explains the asymmetry
seen in many profiles (Han \& Manchester 2001\nocite{hm01}).

Unfortunately there exists no sound theoretical basis for the radio
emission from pulsars. This makes it difficult to tie down the
observed phenomenology of the integrated profiles to any physical
understanding of the conditions in the magnetosphere. However, conal
structures are predicted by some models (e.g. Ruderman \& Sutherland
1975\nocite{rs75}).

We outline in this paper the basic ingredients of an empirical model
for the three dimensional location of radio emitting regions in the
pulsar magnetosphere. The main purpose of this model is to account for
profiles which obey what we consider the current strongest
observational constraints with only a small number of free
parameters. As we are mostly unable to well determine the true viewing
geometry of each observed pulsar, our approach in testing the model is
rather a statistical one. In generating large numbers of artificial
profiles, we consider to have chosen reasonable parameters for our
model when we reproduce a similar range of profile characteristics as
in our observed sample.

In the following section we review the important observational results
related to integrated pulse profiles. We then outline our model and
discuss its parameters. We show how the results from simulations match
the observational data, draw some initial conclusions and outline our
future work.

\section{Observational background}
\subsection{Pulsar phenomenology}
Pulsar radio emission is thought to arise from the open field lines,
with angular radius $\rho$ surrounding the magnetic pole. The pole is
inclined by an angle $\alpha$ with respect to the rotation axis. The
line of sight traverses the beam; the closest approach of the line of
sight to the magnetic axis is the angle $\beta$.  Under this geometry,
the observed pulse width, $W$, is related to $\rho$ via
\begin{equation}
\sin^2\left(\frac{W}{4}\right)
=\frac{\sin^2(\rho/2)-\sin^2(\beta/2)}{\sin\alpha\cdot\sin(\alpha+\beta)}
\end{equation}
(Gil et al. 1984)\nocite{ggr84}.  For regions close to the magnetic
axis and dipolar field lines, $\rho$ is related to the height $H$
above the surface of the star:
\begin{equation}
\rho \sim \sqrt{\frac{9\pi H}{2cP}},
\end{equation}
where $c$ is the speed of light and $P$ the pulse period.

In explaining the shape of pulsar profiles, there are several
observational lines of evidence which we believe any phenomenological
model should reproduce. These are
\begin{itemize}

\item {\bf single component profiles:} A large fraction of the
observed pulsar profiles consist of a single component. Increased
temporal resolution may, in many cases, indicate that the single
component consists of several Gaussian components that largely
overlap (e.g. Wu et al. 1998)\nocite{wgr+98}.

\item {\bf evidence for conal emission:} In the profiles that are not
single, symmetry is often present (e.g. PSRs B1133+16 and B0525+21).
In such cases the most natural explanation is that the emission 
originates from a ring centered on the magnetic pole.

\item {\bf the period-width relationship:} Rankin has shown good
evidence that the profile width, $W$, depends on the pulsar period,
$P$, as $W\propto P^{-0.5}$. This is an important result and implies
(see equation 2) that the height at which a pulsar emits is 
largely independent of its period (Rankin 1993)\nocite{ran93}.

\item {\bf the emission height versus pulse longitude:} More than two
decades ago, Krishnamohan \& Downs (1983)\nocite{kd83} put together an
attractive cartoon to explain the properties of the Vela pulsar by
proposing that emission components originated from different heights
in the magnetosphere at a particular frequency.  Rankin
(1993)\nocite{ran93} dwells briefly on this idea as an alternative
explanation to nested cones.  More recently, Gangadhara \& Gupta
(2001)\nocite{gg01} showed for PSR B0329+54 that the emission heights
at a given frequency vary as a function of pulse longitude with high
emission heights seen at the edges of profiles and low emission
heights in the central parts of the profile.  Furthermore, Mitra \&
Rankin (2002)\nocite{mr02a} show that inner cones originate from lower
in the magnetosphere than outer cones.  The combination of these
results is a key result which breaks the long-held assumption that
emission at a given frequency arises from a uniform height above the
pole.

\item {\bf the simplicity of young pulsar profiles:} Johnston \&
Weisberg (2006)\nocite{jw06} showed that the profiles of a group of
young pulsars can be reproduced simply by postulating a single, rather
wide, cone of emission located relatively high in the magnetosphere
following an earlier idea by Manchester (1996)\nocite{man96}. This is
in contrast to older pulsars where complex, multi-component profiles
are often observed. Differences between young and old pulsars are not
only related to the shape of the profile and the spin-down energy, but
also to the increased timing noise seen in young pulsars, as opposed
to older pulsars. The transition from one group to the other has not
yet been explored in this respect.

\item {\bf that cones are not fully illuminated:} The Lyne \&
Manchester (1988)\nocite{lm88} model for pulsar beams, which takes
into consideration a large amount of observational data available at
that time, suggests that the emission within the beam boundary occurs
in patches with essentially random locations. Occasionally these
sub-beams can form regular patterns such as those seen in drifting
sub-pulses (Backer 1973)\nocite{bac73}.

\item {\bf that complex profiles have complex position angle (PA)
swings:} It is very noticeable that many simple conal profiles also
have the PA swings expected in the rotating-vector-model (RVM, where
the PA is fixed by the plane of curvature of the emitting magnetic field
line). On the other hand, complex profiles often show strong
deviations from the standard RVM picture. The natural explanation here
is that overlapping components at different heights cause a distortion
of the observed PA swing (Karastergiou \& Johnston 2006)\nocite{kj06}.

\item {\bf that profiles get wider with decreasing observing
frequency:} The so-called radius to frequency mapping (RFM) idea is
that lower frequencies are emitted higher in the magnetosphere than
higher frequencies which naturally results in wider observed profiles
(see equation 2). Thorsett (1991)\nocite{tho91a} postulated a
width-frequency law which essentially accounts for the fact that pulse
widening only really occurs at frequencies below $\sim$1~GHz; above
this value the pulse width is constant.
\end{itemize}
\begin{table}
\caption{Profile classification for 283 pulsars}
\begin{tabular}{lcccc}
\hline & \vspace{-3mm} \\
log$\dot{E}$ & Single & Double & Multiple & Total\\
\hline & \vspace{-3mm} \\
$<$ 35.0&    54\% &    38\%  &   8\%  &     26\\
33-35   &  47\%   &  23\%  &  30\%    &   93\\
32-33   &  48\%   &  22\%  &  30\%    &   74\\
$<$ 32  &     49\% &    21\% &   30\% &      90\\
\hline & \vspace{-3mm} \\
log($\tau_c$) & & & & \\
\hline & \vspace{-3mm} \\
$<$5.0 &      58\% &     38\% &     4\% &       24\\
5-6.5   &  47\% &     23\% &     30\% &      98\\
6.5-7.0 &  37\% &     30\% &     33\% &      60\\
$>$7     &   55\% &     16\% &     29\% &      101\\
\hline & \vspace{-3mm} \\
period [ms] & & & & \\
\hline & \vspace{-3mm} \\
$<$150    &  61\% &     39\% &      0\% &      18\\
150-400 &  46\% &     23\% &     31\% &      106\\
400-700 &  51\% &     19\% &     30\% &      82\\
$>$700    &  46\% &     26\% &     28\% &      77\\
\end{tabular}
\label{class}
\end{table}
\begin{figure*}
\begin{center}
\resizebox{0.8\hsize}{!}{\includegraphics[angle=0]{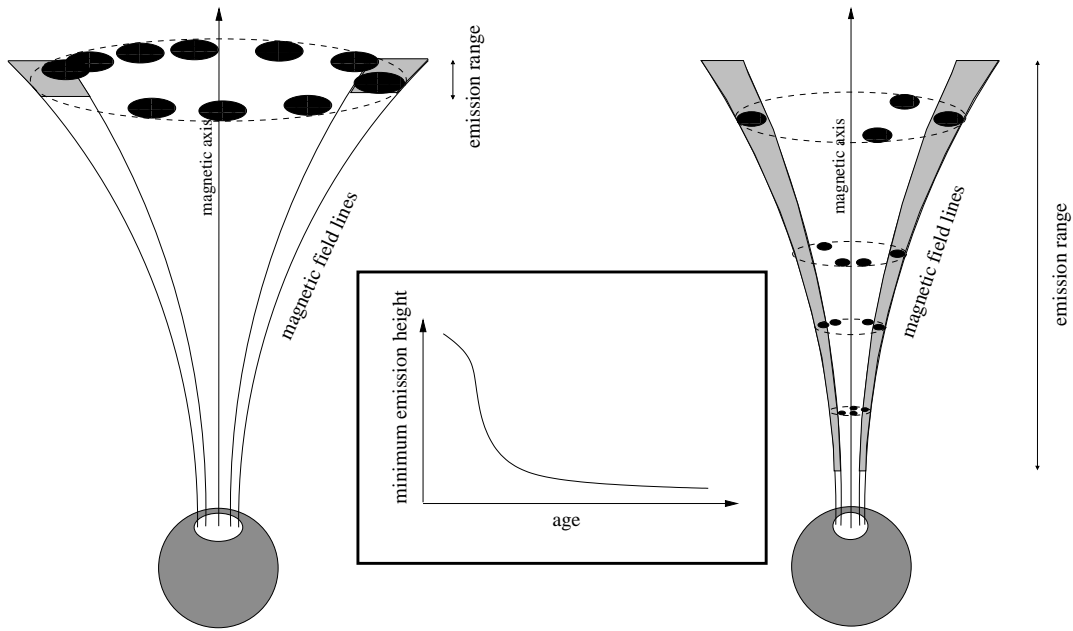}}
\end{center}
\caption{A cartoon of the proposed model. On the left we show the
model for young pulsars with emission arising from a patchy conal ring
over a narrow range at high altitudes. On the right, the case for
older pulsars is shown. Here, the polar cap is smaller and emission at
a given frequency arises from close to the surface to a height of
$\sim$1000~km in a series of discrete patchy rings. The middle panel
shows the proposed variation in the minimum emission height with
increasing characteristic age, increasing period and decreasing
$\dot{E}$}.
\label{f1}
\end{figure*}
\subsection{Profile classification}
We have recently obtained polarization data on more than 250 pulsars,
using an unbiased sample of pulsars in the southern sky strong enough
to give high signal to noise profiles in less than 30 minutes at the
Parkes radio telescope. Our aim was to establish a new, large database
of pulsar profiles observed at multiple frequencies with high temporal
resolution (Karastergiou et al. 2005; Karastergiou \& Johnston 2006;
Johnston et al. 2006)\nocite{kjm05,jkw06}.  We devised a simple
classification scheme by visual inspection of the profiles. The
temporal resolution of the profiles was large, with up to 2048 bins
per profile.  Profiles were classified into `single', `double' and
`multiple' component profiles, and then sorted by their spin down
energy, $\dot{E}$, characteristic age, $\tau_c=P/(2\dot{P})$, and
period, $P$. The results are listed in Table~\ref{class}.

A priori, it is not clear how the complexity of the pulse profile
relates back to physical parameters. The spin down energy determines
the power available but the radio luminosity is only a tiny fraction
of this power. The pulse period affects the size of the polar cap and
light cylinder and should likely be important. The age, on the other
hand, is hard to determine with accuracy and may not play a direct
role in the energetics.

In Table~\ref{class}, however, we see a very striking result which is
replicated in all three parameters. The young, fast-spinning, highly
energetic pulsars {\bf do not} have complex profiles and have roughly
a 60:40 split between single and double component profiles.  In
contrast, the older, slower spinning, less energetic stars do have
complex profiles and the split between single, double and mulitple
component pulsars is roughly 45:25:30. Furthermore, there appears to
be a rather abrupt transition where complex profiles are formed at
$P\sim150$~ms, $\tau_c\sim10^5$~yr and/or
$\dot{E}\sim10^{35}$~ergs$^{-1}$.  Above this transition point, the
relative fractions stay constant. In the following, when we refer to
young pulsars, we mean young/short-period/highly-energetic pulsars below this
transition point, as opposed to older, slower and less energetic
pulsars. The distinguishing parameter we use in the simulations is the
pulse period.

\section{A simple beam model}
We now attempt to understand the observational evidence and reconcile
the nested cone models of Rankin with the patchy models of Lyne \&
Manchester in a simple way.  We postulate a beam model which has the
following ingredients:
\begin{itemize}
\item radio emission originates from field lines close to the outer
edge of the beam, forming an emission cone;
\item the conal beam is patchy;
\item the maximum altitude of emission, in all pulsars, is set to
$\sim$1000~km at a frequency of $\sim$1~GHz;
\item the minimum altitude of emission is large for young pulsars
  (similar to the maximum altitude) and small ($\sim$20~km) for older
  pulsars;
\item emission arising from discrete locations within the entire range
of emission heights is possible at a given frequency;
\item the polarization PA of each patch is tied to the magnetic field
line at the centre of that patch.
\end{itemize}

Figure~\ref{f1} illustrates the model which naturally reproduces
virtually all the observational phenomenology outlined in section
2. In particular, in older pulsars, the large range of allowable
heights gives rise to apparent ``nested'' emission zones resulting in a
complex profile.

The limitations we place on the origins of emission within the radio
beam immediately pose a question as to the nature of so-called 'core'
emission in pulsars. In our approach here, central components, which
are often seen to have special properties (for example a steeper
spectrum), are tangential cuts of emission patches which are deeper in
the beam structure and therefore naturally closer to the magnetic
axis. As these components arise from lower altitudes, they are
naturally flanked by components originating from higher up in the
magnetosphere. In a similar geometrical approach, Sieber (1997)
attempted an explanation of 'core' components within the framework of
the nested cone model. 

\begin{figure*}
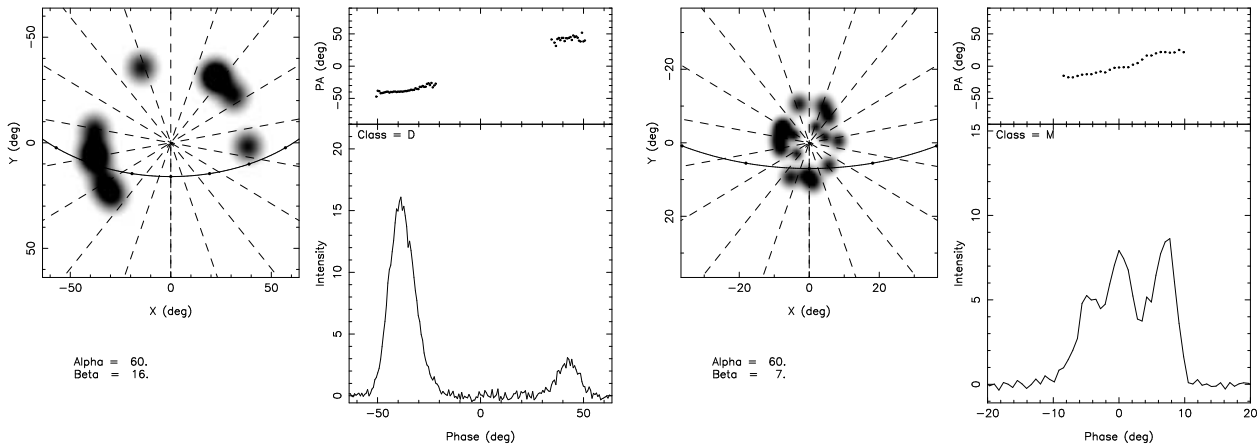

\begin{tabular}{cc}
\includegraphics[width=5.8cm,angle=-90]{young.ps} &
\includegraphics[width=5.8cm,angle=-90]{old.ps}\\
\end{tabular}
\caption{Artificial beams and profiles generated for a young pulsar
with period of 100~ms (left) and an older pulsar with period of
1000~ms (right). The beam is shown from above the magnetic pole, and
the line of sight is also depicted. Note the narrower patches
originating from lower altitudes in the beam depicted on the
right. Above the pulse profile, we show the polarization PA obtained
according to the model. Deviations from the simple RVM are caused by
overlapping patches. Longitude zero depicts the magnetic pole
crossing. }
\label{examples}
\end{figure*}
In the following, we assume that all emitting patches within the beam
have the same peak brightness. We attribute profiles with components
of different strengths to the specific line-of-sight cut and
overlapping components (see Figure~\ref{examples}). Observationally,
this seems reasonable, as it is rare that pulsar profiles contain
either very large or very tiny amplitude components. At the same time,
pulsars are not standard candles, but the reason for a given pulsar's
luminosity is not known, nor does there appear to be any strong
correlation between luminosity and any other pulsar parameter. Also,
as we are postulating patches of emission, we create inhomogeneity in
the brightness of the pulsar beam. There is no way that this
brightness distribution can be decomposed uniquely into patches of
variable brightness.

Out of the model ingredients, which determine the three-dimensional
beam, some are well constrained by observations and can be fixed to
certain values, while others are unconstrained and explored using
numerical simulations. For the minimum and maximum emission heights,
and the size of the emitting patches, there are a number of
assumptions which guide us to the values used in our simulations,
which we explain in more detail in the following section. On the other
hand, we have no previous constraints on the discrete emitting heights
and patches of emission per height, and are forced to explore a large
area of the parameter space to appreciate their role.

\subsection{Constraints to the model}
As mentioned earlier, Rankin (1993) shows convincing evidence that the
maximum emission height is independent of the pulsar period. Typical
maximum emission heights from the literature range from $\sim$10 to
$\sim$1000~km above the stellar surface (Blaskiewicz et al. 1991,
Mitra \& Li 2004\nocite{bcw91,ml04}).  The model presented here
retains a constant maximum height for pulsars of all periods. For our
simulations, we fixed the maximum emission height at 1000~km at a
frequency of 1~GHz.

For the minimum height, we consider two populations distinguished by
the period $P$. In order to reproduce the values shown in
Table~\ref{class} we set the minimum height to be very near the
maximum height in short period pulsars ($P<0.15$~s).  For longer
periods, the higher complexity in the pulse profiles leads us to set
the minimum height much further down in the magnetosphere. In our
simulations, we choose a minimum height just above the stellar surface
for all pulsars with $P>0.15$~s. We set this abrupt transition in
accordance with the profile classification in section 2.2.

In this context, it is interesting to examine the consequences of
setting the emission height range to be a constant fraction of the
neutron star light cylinder radius. For example, a pulsar with 0.1~s
period, has a light cylinder radius of 4800~km. If radio emission
occurred over a height range spanning 2\% of the light cylinder
radius, this corresponds to $\sim$100~km and so the allowable emission
heights would range from 900 to 1000~km. For an older, 1~s period
pulsar, the light cylinder grows to 48000~km, 2\% now corresponds to
$\sim$1000~km and the entire range from the surface of the star to the
maximum emission height is now allowed. Although this line of thinking
is appealing, it presents difficulties in explaining the abrupt change
in the profile classification seen in Table~\ref{class} and discussed
earlier.

The width of a patch of emission at a given height is a key parameter
for the model.  We use two sources to derive this. First, Johnston \&
Weisberg (2006) estimate that the conal width, $\Delta s$ in young
pulsars is about 20\% of the total width available. Secondly, Mitra \&
Rankin (2002) measure component widths as a function of emission
height in a variety of double component pulsars. We use their equation
to assign a patch width based on emission height
\begin{equation}
w_p = 2.45\degr \Delta s \sqrt{\frac{H}{10\cdot P}},
\end{equation}
with $\Delta s=0.2$ and $H$ expressed in km.
\begin{figure*}
\centerline{
\includegraphics[width=16cm]{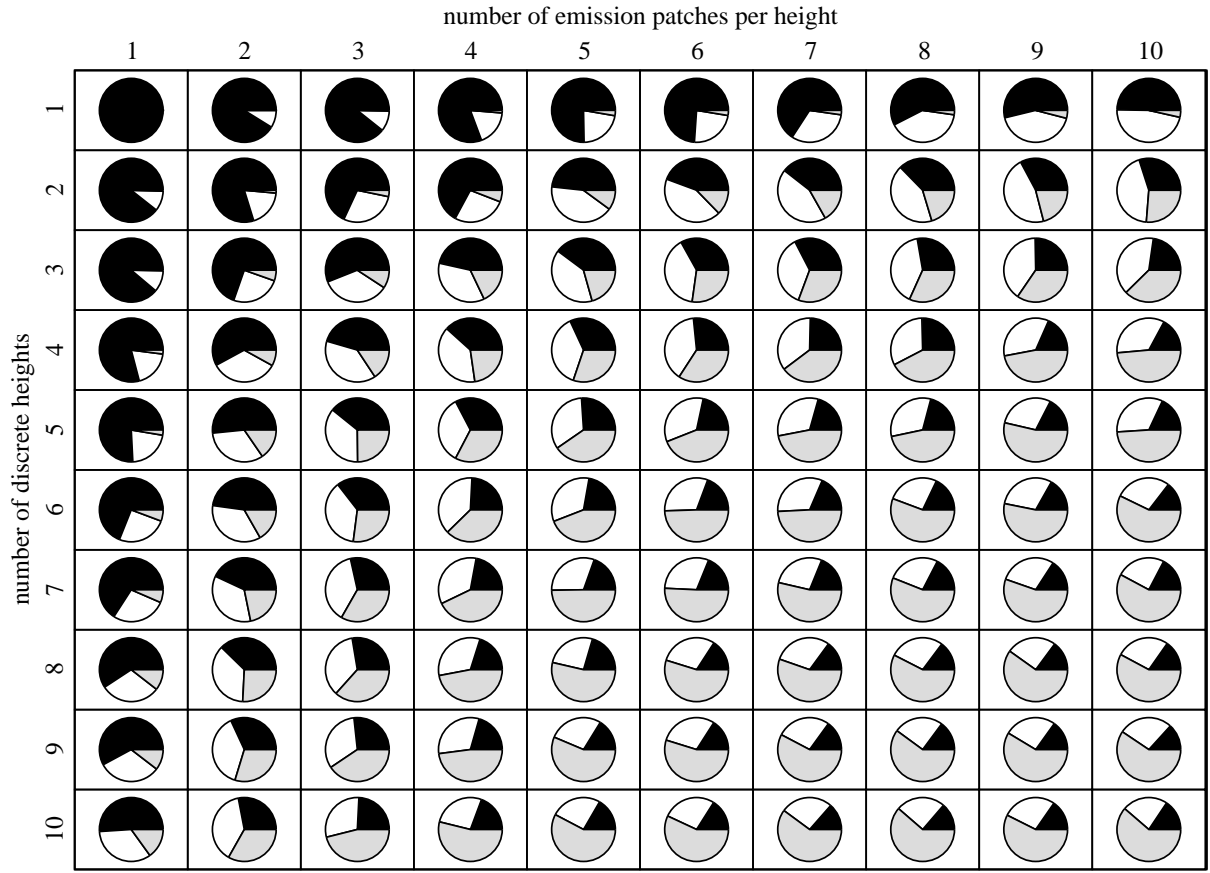}
}
\caption{The results of a numerical simulation of 10$^5$ average
  profiles for a pulse period of 0.3~s. $10^3$ artificial profiles are
  generated for each combination of number of heights (y-axis) and
  number of patches (x-axis). The pie charts show the fraction of
  generated profiles which are single- (black), double- (white) and
  multiple-component (grey). }
\label{10by10}
\end{figure*}
\section{Results}
A large number of numerical simulations have been performed to
constrain the remaining free parameters, namely the active altitudes
and the patches per emission height. As concerns the number of heights
and patches, we have explored the parameter space ranging from 1 to 10
discrete heights and 1 to 10 patches per height. We have no a priori
knowledge on the location of these heights and patches and pick their
positions randomly in our simulations. These consist of ``creating''
patchy beams according to the model. Each beam is then systematically
cut with 300 different lines of sight, spanning a reasonable range of
$\alpha$ and $\beta$. This results in a ``clean'' profile for each
line of sight, to which an amount of Gaussian noise is applied before
passing it through an automated classification process to determine
whether the generated profile is single, double or multiple. This
process takes the local maxima before the addition of noise and
assesses them after noise addition to determine whether their
separation in the intensity-time diagram is adequate compared to the
noise. Empirical fine tuning was applied to match the results with our
own visual interpretation of the profiles. The process was repeated a
large number of times due to the randomness in the discrete heights
and location of the patches. Following this procedure, we check what
fraction of the artificially generated profiles are single, double or
multiple for a given combination of emitting heights and patches per
height. This test also results in a predicted beaming fraction, that
is the fraction of pulsars we would expect to detect given a certain
beam configuration.

In a second test, we create large sets of 1000 average profiles per
combination of emitting heights and patches per height, using a
randomly computed viewing geometry per profile. We chose from a subset
of $\alpha$ and $\beta$ which guarantees that the line of sight
intersects the cone of emission. Figure~\ref{10by10} shows an example
of such a simulation of $10\times10\times1000$ beams. The pie chart
within each box represents the relative fraction of single component
profiles (black), double component profiles (white) and multiple
component profiles (grey), simulated with a period of 0.3~s. The plot
demonstrates clearly how a single emitting altitude (top line) can not
account for a substantial fraction of multiple component profiles. It
also demonstrates how, as the total number of heights and components
grows, the fraction of single profiles becomes small.

\subsection{Young, short-period pulsars}
The observational data suggests that over 50\% of the profiles in this
category comprise of a single component. The rest are almost all
symmetrical doubles and only a very small number of multiple component
profiles exist (see Table~1).  Evidence that emission from such
objects generally arises from high altitudes has been presented in
Johnston \& Weisberg (2006) and we therefore limit our range of
emission heights from 950 to 1000~km. However, the fraction of single
profiles also forces the total number of patches to be rather small,
so that the line of sight will intersect only one patch on half of the
observed cases. We find the best value to be 10 patches in total, at a
single emission height. Our simulations also provide a rough estimate
of the beaming fraction. For this group, we find the beaming fraction
to be around 30\%. An example of a simulated artificial beam and
profile from this group can be seen in the left hand panel of
Figure~\ref{examples}.

\begin{figure*}
\begin{tabular}{cc}
\includegraphics[width=6.8cm,angle=-90]{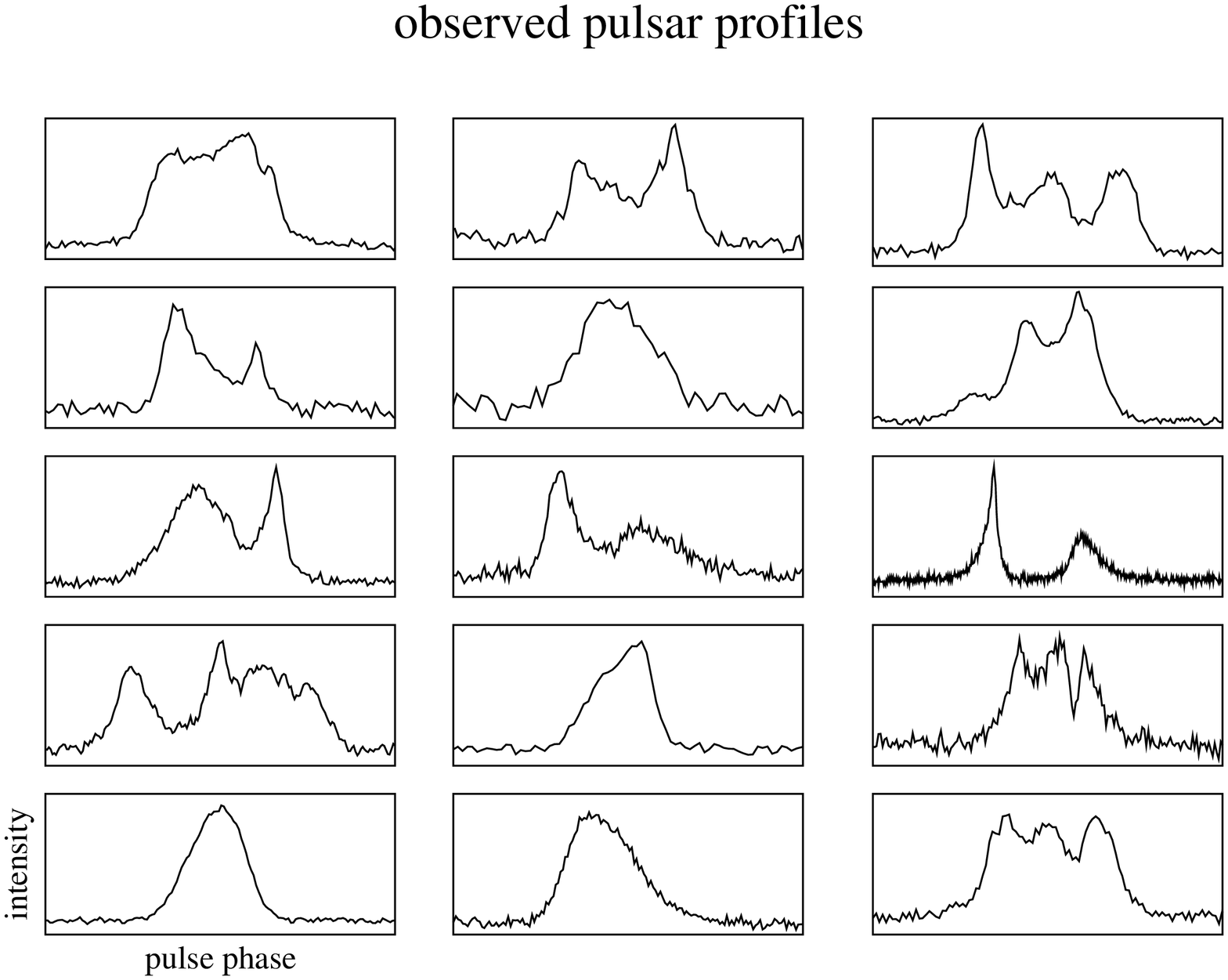} &
\includegraphics[width=6.8cm,angle=-90]{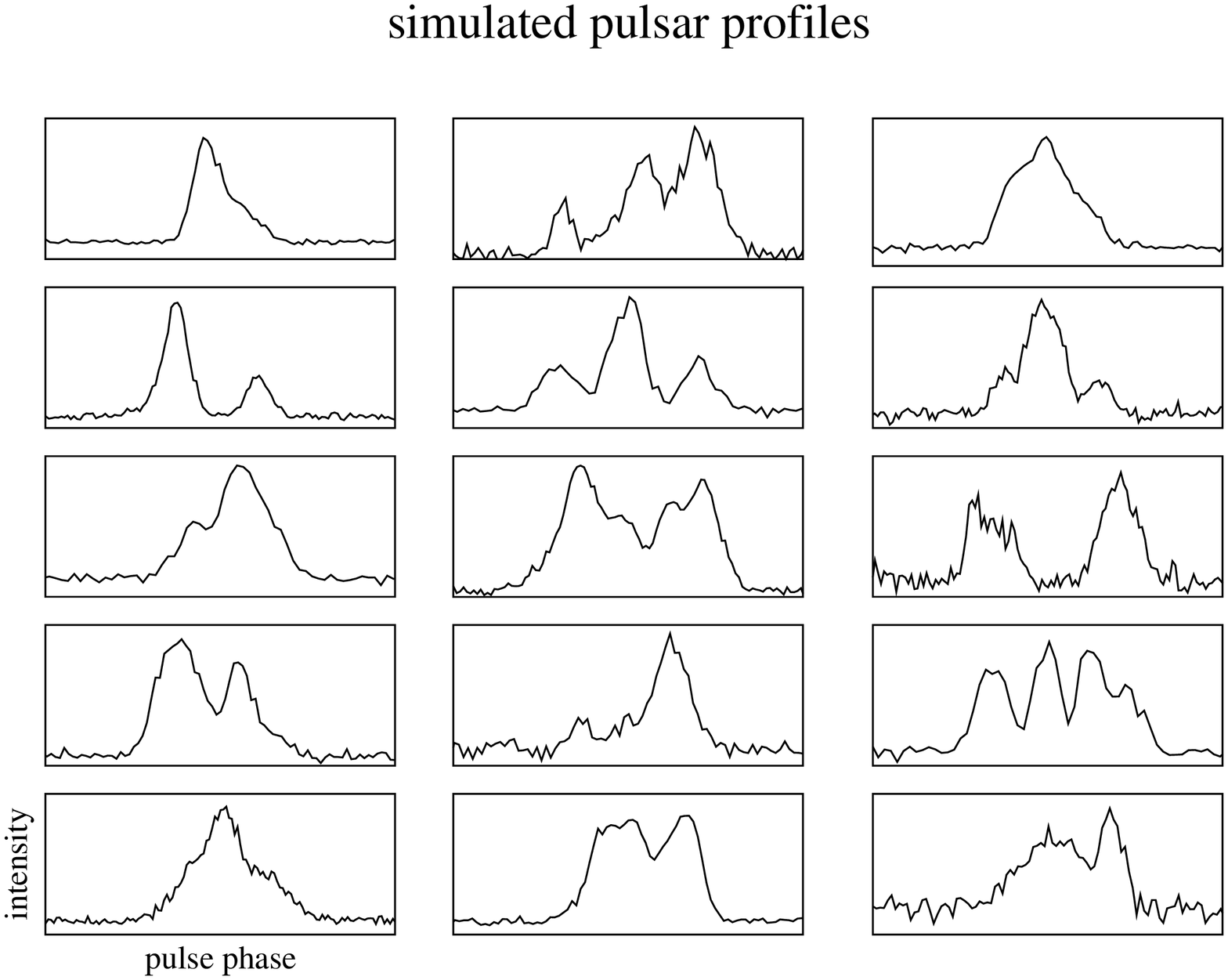}\\
\end{tabular}
\caption{The left panel shows fifteen profiles randomly selected from
our database of pulsar polarization data obtained at 1.4~GHz with the
Parkes radio telescope. These show the usual variety of phenomenology
such as multiple components and components of various width,
amplitude and shape.  On the right are fifteen simulated profiles according to
our model.  Although a direct one-to-one comparison is not our
intention here, the same sort of profile variety as exists in the real
data is clearly reproduced.  }
\label{random}
\end{figure*}
\subsection{Older, slower pulsars}
In contrast to young pulsars, this group shows large diversity in the
shapes of the observed profiles. In particular there is a large
fraction of multiple component profiles (see Table~1) but this
fraction does not appear to change either with $\dot{E}$, $tau_c$ or
$P$.  In our model, we attribute this diversity to the wide range of
active emission heights at a given frequency.

We note that as a pulsar's period increases, because the maximum
emission height is constant, both $\rho$ and $W$ must decrease (see
equations 1 and 2). At the same time individual component widths
decrease in a similar fashion according to equation 3. This naturally
forces pulse profiles to be very similar looking (although with
different widths) for all slower pulsars, similar to the statistics
from Table~1.  We find that we can reproduce these statistics as long
as we permit at least 3 active heights, and a number of active patches
per height roughly ranging from 2 to 7, depending on the number of
heights. We obtain the best match to the observations when the total
number of patches is $\sim$16$\pm4$ (example, for 4 active heights,
3-5 active patches per height). Pulsars with periods of 0.2~s have a
beaming fraction of around $\sim$18\%, whereas only $\sim$5\% of 2~s
pulsars would be detected, similar to results obtained from population
studies (Tauris \& Manchester 1998)\nocite{tm98}. An example of a
simulated artificial beam and profile from this group can be seen in
the right hand panel of Figure~\ref{examples}.

A visual inspection of the artificial profiles yields a strong affirmation
of a hollow cone of emission, with many symmetrical double profiles. A
number of instances of characteristic triple profiles, with a strong
central component and weaker outrider components also randomly arise.
We are also encouraged by the fact that our simulations never result
in profiles with an unacceptably high number of distinct components,
with a maximum of $\sim$5 similar to the observed population. To
illustrate these points, Figure~\ref{random} shows a comparison
between randomly selected, artificial pulse profiles and real observational
data.
\subsection{Radius-to-frequency mapping - RFM}
\begin{figure*}
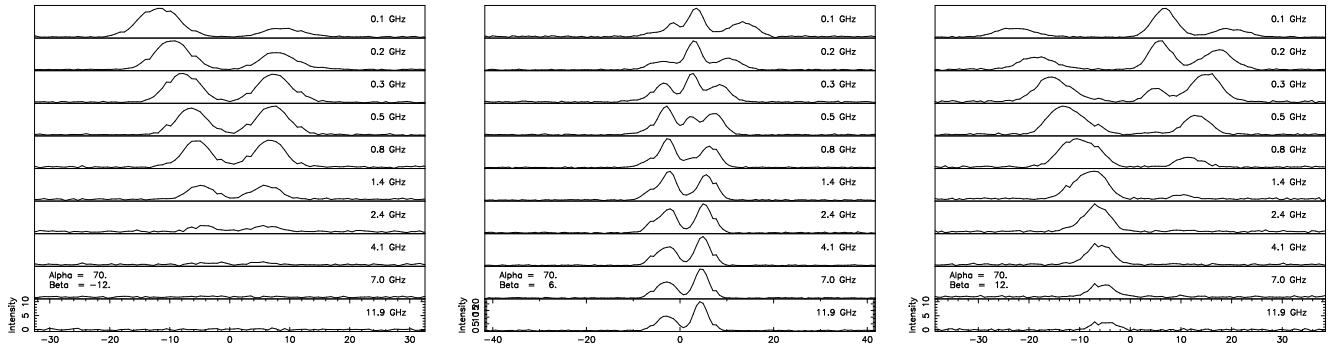

\begin{tabular}{ccc}
\includegraphics[width=4.5cm,angle=-90]{rfm1.ps} &
\includegraphics[width=4.5cm,angle=-90]{rfm2.ps} &
\includegraphics[width=4.5cm,angle=-90]{rfm3.ps}\\
\end{tabular}
\caption{Three examples of simulated profiles at 10 different
observing frequencies. As a consequence of equation~4, the pulse
profile becomes narrower at higher observing frequencies.  In all
cases there are substantial profile shape changes between low and high
frequencies. The middle and right hand panels show that the smaller,
lower altitude patches seen towards the centre of the profile have an
apparently steeper spectral index than the outrider components,
similar to what is observed in the known pulsars.  The panel on the
left demonstrates how geometry hampers high frequency observations, as
the line of sight no longer intersects the pulsar beam.}
\label{rfm}
\end{figure*}

Up to this point, we have drawn results from simulations at a single
observing frequency near 1~GHz.  However, we can also add RFM to our
model, following ideas from Rankin \& Mitra (2002). They interpreted
the earlier work of Thorsett (1991)\nocite{tho91a} and showed that the
height of emission has a functional form
\begin{equation}
H_{\nu}=K\cdot\nu^{-2/3}+H_{0},
\end{equation}
with $K$ and $H_{0}$ picked to ensure little evolution of the profile
width above $\sim$1~GHz. In the standard model, $H_{\nu}$ is constant
across the pulse. In our model, however, we compute $H_{\nu}$ for
every created patch.  Figure~\ref{rfm} shows three examples obtained for
our implementation of RFM. The impact angle $\beta$ changes from left
to right, from $-6\degr$ to $6\degr$ to $12\degr$. The profile width
changes with frequency according to equation 4 combined with equation
2.  Note the apparent change of the component amplitude ratios as a
function of frequency. This is especially striking in the second and
third examples where a central component at low frequencies quickly
disappears at high frequencies. This apparent spectral index change
between components is then purely geometric in this model as
originally suggested by Sieber (1997)\nocite{sie97}.  In the first
panel of Figure~\ref{rfm}, the pulse profile becomes very weak above
4~GHz as the beam becomes narrower and is no longer directed towards
the line-of-sight.

\subsection{The polarization position angle}
In this paper
we do not intend to present a detailed treatment of the polarization of
average pulsar profiles in the context of our model. However, it is useful to
look at the consequences of our model, particularly on the PA
swing.  In particular, we note the disruption of smooth average PA
swings in complex profiles and believe our model can also reproduce this. In
the model, we tie the PA of each patch to the magnetic field line
of the centre of the patch. In complex profiles, where there may be several
overlapping patches, this causes significant distortion of the (geometrical)
PA angle swing which is qualitatively similar to real data (see
Figure~\ref{examples}).
\section{Summary and future work}

We have developed a beam model for the radio emission from pulsars,
where emission is generated at more than one discrete height at a
given frequency. The emission occurs in a ring close to the last open
field lines. Emission from a given ring is patchy. Following extensive
numerical simulations, Table~\ref{pars} shows the parameters from the
model which best reproduce the features of the observational data. The
model yields
\begin{itemize}
\item the simple profiles of young, energetic pulsars as opposed to
  more complex profiles of the older population;
\item a wide variety of simulated profiles which closely resembles the
  observations;
\item an explanation for the number of single-component profiles 
      through the patchiness of the ring structure at a given height;
\item a complex PA angle profile for complex profiles,
  as in the observed data;
\item RFM behaviour consistent with the observations;
\item beaming fractions as a function of period 
  which are consistent with other studies.
\end{itemize}

From a theoretical point of view, the idea that emission arises only
from near the last open field lines is appealing. Our model requires a
successful theory that gives rise to emission at many heights at a
particular frequency rather than emission at many locations across the
polar gap. As the emission properties are tied somehow to the plasma
conditions, one could postulate a varying plasma density along the
vertical extent of the emission tube as a possible explanation for
multiple emission heights.
\begin{table}
\caption{Model Parameters}
\begin{tabular}{l|ll}
\hline & \vspace{-3mm} \\
Pulsar group & Young & Old \\
\hline & \vspace{-3mm} \\
Maximum height: & 1000~km & 1000~km \\
Minimum height: & 950~km & 20~km \\
Distinct emission heights: & 1 & $>3$ \\
Active patches per height: & 10 & 2 ... 7 \\
Total number of patches: & $10\pm1$ & $16\pm4$ \\
\hline & \vspace{-3mm} \\
\end{tabular}
\label{pars}
\end{table}

We recognise that there are several observational differences between
the outer and inner parts of pulsar profiles. In particular they may
have different drift rates and certainly have different modulation
properties. Generally this has been assumed to be caused by different
emission properties between the outer and inner magnetic field lines.
However, in the absence of any physical model, this could equally
likely be caused by differences between higher and lower altitudes as
would be the case in our model. A lot of important work is being
carried out presently on the drifting properties of individual pulses
(e.g. Weltevrede et al. 2006, 2007)\nocite{wes06,wse07}, some of it
indicating close relationships between the drift rates of the various
rings that are postulated to form the beam (Bhattacharyya et
al. 2007)\nocite{bgg+07}, although a clear picture has yet to emerge
concerning how these drift patterns ultimately form an average pulse
profile.

Currently our model is designed to reproduce the total intensity
profiles and we therefore only lightly touch on polarization aspects
of pulsar emission. Much exciting work has also been done in this area
recently and we plan to incorporate polarization into the next
iteration of the model.

In summary we have produced a simple beam model of pulsar radio
emission, which can generate average pulse profiles in accordance with
the most current observational constraints. Its main features are that
it postulates emission over a wide range of emission heights rather
than over a wide range of beam longitudes as in previous models and
that it largely replicates the observational data.

\section*{Acknowledgments}
We thank Don Melrose and Steve Ord for valuable discussions during the
writing of this paper and Michael Kramer for parts of the simulations
code.  AK particularly thanks Don Melrose at the University of Sydney
for his hospitality and acknowledges financial support from the 6th
European Community Framework programme through a Marie Curie,
Intra-European Fellowship. The Australia Telescope is funded by the
Commonwealth of Australia for operation as a National Facility managed
by the CSIRO.

\bibliography{journals,modrefs,psrrefs,crossrefs} 

\begin{thebibliography}{}

\bibitem[\protect\citeauthoryear{Backer}{Backer}{1973}]{bac73}
Backer D.~C.,  1973, ApJ, 182, 245

\bibitem[\protect\citeauthoryear{Bhattacharyya, Gupta, J. Gil \&
M. Sendyk}{Bhattacharyya et~al.}{2007}]{bgg+07} Bhattacharyya B.,
Gupta Y., Gil J., Sendyk M., 2007, MNRAS, 377, L10

\bibitem[\protect\citeauthoryear{Blaskiewicz, Cordes \&
  Wasserman}{Blaskiewicz et~al.}{1991}]{bcw91} Blaskiewicz M., Cordes
  J.~M., Wasserman I., 1991, ApJ, 370, 643

\bibitem[\protect\citeauthoryear{{Gangadhara} \& {Gupta}}{{Gangadhara} \&
  {Gupta}}{2001}]{gg01}
{Gangadhara} R.~T.,  {Gupta} Y.,  2001, ApJ, 555, 31

\bibitem[\protect\citeauthoryear{Gil, Gronkowski \& Rudnicki}{Gil
  et~al.}{1984}]{ggr84}
Gil J.~A.,  Gronkowski P.,    Rudnicki W.,  1984, A\&A, 132, 312

\bibitem[\protect\citeauthoryear{{Han} \& {Manchester}}{{Han} \&
  {Manchester}}{2001}]{hm01}
{Han} J.~L.,  {Manchester} R.~N.,  2001, MNRAS, 320, L35

\bibitem[\protect\citeauthoryear{{Johnston}, {Karastergiou} \&
  {Willett}}{{Johnston} et~al.}{2006}]{jkw06}
{Johnston} S.,  {Karastergiou} A.,    {Willett} K.,  2006, MNRAS, 369, 1916

\bibitem[\protect\citeauthoryear{{Johnston} \& {Weisberg}}{{Johnston} \&
  {Weisberg}}{2006}]{jw06}
{Johnston} S.,  {Weisberg} J.~M.,  2006, MNRAS, 368, 1856

\bibitem[\protect\citeauthoryear{{Karastergiou} \& {Johnston}}{{Karastergiou}
  \& {Johnston}}{2006}]{kj06}
{Karastergiou} A.,  {Johnston} S.,  2006, MNRAS, 365, 353

\bibitem[\protect\citeauthoryear{{Karastergiou}, {Johnston} \&
  {Manchester}}{{Karastergiou} et~al.}{2005}]{kjm05}
{Karastergiou} A.,  {Johnston} S.,    {Manchester} R.~N.,  2005, MNRAS, 359,
  481

\bibitem[\protect\citeauthoryear{{Kramer}, {Stairs}, {Manchester},
  {McLaughlin}, {Lyne}, {Ferdman}, {Burgay}, {Lorimer}, {Possenti}, {D'Amico},
  {Sarkissian}, {Hobbs}, {Reynolds}, {Freire} \& {Camilo}}{{Kramer}
  et~al.}{2006}]{ksm+06}
{Kramer} M.,  {Stairs} I.~H.,  {Manchester} R.~N.,  {McLaughlin} M.~A.,  {Lyne}
  A.~G.,  {Ferdman} R.~D.,  {Burgay} M.,  {Lorimer} D.~R.,  {Possenti} A.,
  {D'Amico} N.,  {Sarkissian} J.~M.,  {Hobbs} G.~B.,  {Reynolds} J.~E.,
  {Freire} P.~C.~C.,    {Camilo} F.,  2006, Science, 314, 97

\bibitem[\protect\citeauthoryear{Kramer, Wielebinski, Jessner, Gil \&
  Seiradakis}{Kramer et~al.}{1994}]{kwj+94}
Kramer M.,  Wielebinski R.,  Jessner A.,  Gil J.~A.,    Seiradakis J.~H.,
  1994, A\&AS, 107, 515

\bibitem[\protect\citeauthoryear{Krishnamohan \& Downs}{Krishnamohan \&
  Downs}{1983}]{kd83}
Krishnamohan S.,  Downs G.~S.,  1983, ApJ, 265, 372

\bibitem[\protect\citeauthoryear{Lyne \& Manchester}{Lyne \&
  Manchester}{1988}]{lm88}
Lyne A.~G.,  Manchester R.~N.,  1988, MNRAS, 234, 477

\bibitem[\protect\citeauthoryear{Manchester}{Manchester}{}]{man96}
Manchester R.~N.,  Wide beams from young pulsars (or one pole for all).
pp 193--196

\bibitem[\protect\citeauthoryear{{Mitra} \& {Li}}{{Mitra} \& {Li}}{2004}]{ml04}
{Mitra} D.,  {Li} X.~H.,  2004, A\&A, 421, 215

\bibitem[\protect\citeauthoryear{Mitra \& Rankin}{Mitra \&
  Rankin}{2002}]{mr02a}
Mitra D.,  Rankin J.~M.,  2002, ApJ, pp 322--336

\bibitem[\protect\citeauthoryear{Rankin}{Rankin}{1983}]{ran83}
Rankin J.~M.,  1983, ApJ, 274, 333

\bibitem[\protect\citeauthoryear{Rankin}{Rankin}{1993}]{ran93}
Rankin J.~M.,  1993, ApJ, 405, 285

\bibitem[\protect\citeauthoryear{Ruderman \& Sutherland}{Ruderman \&
  Sutherland}{1975}]{rs75}
Ruderman M.~A.,  Sutherland P.~G.,  1975, ApJ, 196, 51

\bibitem[\protect\citeauthoryear{Sieber}{Sieber}{1997}]{sie97}
Sieber W.,  1997, A\&A, 321, 519

\bibitem[\protect\citeauthoryear{Tauris \& Manchester}{Tauris \&
  Manchester}{1998}]{tm98}
Tauris T.~M.,  Manchester R.~N.,  1998, MNRAS, 298, 625

\bibitem[\protect\citeauthoryear{Thorsett}{Thorsett}{1991}]{tho91a}
Thorsett S.~E.,  1991, ApJ, 377, 263

\bibitem[\protect\citeauthoryear{{Weltevrede}, {Edwards} \&
    {Stappers}}{{Weltevrede} et~al.}{2006}]{wes06} 
{Weltevrede} P., {Edwards} R.~T., {Stappers} B.~W., 2006, A\&A, 445, 243

\bibitem[\protect\citeauthoryear{{Weltevrede}, {Stappers} \&
    {Edwards}}{{Weltevrede} et~al.}{2006}]{wse07} 
{Weltevrede} P., {Stappers} B.~W., {Edwards} R.~T., 2007, A\&A, 469, 607

\bibitem[\protect\citeauthoryear{{Wu}, {Gao}, {Rankin}, {Xu} \&
  {Malofeev}}{{Wu} et~al.}{1998}]{wgr+98}
{Wu} X.,  {Gao} X.,  {Rankin} J.~M.,  {Xu} W.,    {Malofeev} V.~M.,  1998,
  Astron. J., 116, 1984

\end{thebibliography}
\bibliographystyle{mn2e}
\label{lastpage} 
\end{document}